\documentclass[aps,twocolumn,showpacs,amsmath,amssymb,prc,superscriptaddress,nofootinbib]{revtex4-1}
\pdfoutput=1 
\usepackage{graphicx,color}
\usepackage{bm}                      
\usepackage{hyperref}               
\usepackage{multirow}

\def \G{\mathcal{G}}
\def \A{\mathcal{A}}
\def \w{\omega} 

\begin{document}

\title{Density and isospin-asymmetry dependence of high-momentum components}

\author{A. Rios}
\affiliation{Department of Physics, Faculty of Engineering and Physical Sciences, 
University of Surrey, 
Guildford, Surrey GU2 7XH, United Kingdom}
 \email{a.rios@surrey.ac.uk}

\author{A. Polls}
\affiliation{Departament d'Estructura i Constituents de la Mat\`eria,
Universitat de Barcelona, E-08028 Barcelona, Spain}

\author{W. H. Dickhoff}
\affiliation{Department of Physics, Washington University, 
St. Louis, Missouri 63130, USA}

\date{\today}

\begin{abstract}
We study the one-body momentum distribution at different densities in nuclear matter, with special emphasis on its components at high momentum. Explicit calculations for finite neutron-proton asymmetry, based on the ladder self-consistent Green's function approach, allow us to access the isospin dependence of momentum distributions and elucidate their role in neutron-rich systems. Comparisons with the deuteron momentum distribution indicate that a substantial proportion of high-momentum components are dominated by tensor correlations. We identify the density dependence of  these tensor correlations in the momentum distributions. Further, we find that high-momentum components are determined by the density of each sub-species and we provide a new isospin asymmetry scaling of these components. We use different realistic nucleon-nucleon interactions to quantify the model dependence of our results.
\end{abstract}



\maketitle

\section{Introduction}

Short-range correlations (SRCs) have been unambiguously identified in a variety of nuclear 
and hadronic physics experiments \cite{Onderwater98,Starink00,Rohe2004,Piasetzky2006,Benhar2008,Subedi2008,Arrington2011,Boeglin2011}. 
Their presence is subtle at the one-body level because they occur at large missing energy and not near 
the Fermi energy~\cite{Rohe2004}. 
At the two-body level, these correlations must occur at all energy scales on account of the expected reduction of the 
amplitude for nucleons to be found at small relative distances.
Ultimately, these two-body correlations are also responsible for the high-momentum components 
of the single-particle distribution.
The quantitative structure of the neutron and proton high-momentum components can now be 
accessed experimentally \cite{Piasetzky2006,Subedi2008,Arrington2011}. 
A clear manifestation of these correlations is the population of high-momentum components of 
the single-particle momentum distribution \cite{Amado1976,Muther1995}. 
The existence of such components, in turn, involves a necessary depletion of 
the low-momentum 
single-particle strength to conserve the total density~\cite{Rios2009a}. 
Green's function techniques are ideally suited to study such effects, as they can incorporate 
SRCs via ladder summation techniques in a self-consistent Green's function (SCGF)   description 
\cite{Dickhoff2008}, including a complete treatment of off-shell effects. 
Moreover, many-body 
approximations can be formulated to preserve basic conservation laws, including the density normalization \cite{Dickhoff2004a}.

Experiments indicate that two-body correlations are dominated by
neutron-proton components \cite{Subedi2008}. 
This suggests that tensor correlations, in addition to SRCs, can play a role in the high-momentum
structure of the nuclear wave function \cite{Schiavilla2007}.
Tensor and SRC act in somewhat different momentum regions, however. A distinction between the 
two is presumably possible if suitable momentum cuts are applied. 
One expects that a sizable population in the neutron-proton (np)
two-body density matrix translates into a relatively large population in the one-body 
momentum distribution. Theoretical calculations of the latter are challenging, as one needs to treat 
and model nuclear short-range components within a many-body environment. 

Revisiting theoretical descriptions of high-momentum components is 
particularly timely, in view of the recently found correlation between the EMC effect and SRC
\cite{Weinstein2011,Arrington2012,Hen2012}. Data are now available for a relatively
large selection of nuclei, which has allowed for the empirical determination of 
the density dependence of parameters that quantify SRC. 
Whereas there is still a discussion regarding two-body center-of-mass
(and other) corrections in establishing the connection between theory and experiment 
\cite{Vanhalst2012,Arrington2012}, the qualitative guidance generated by such experiments is very valuable. 
We discuss the density dependence of such correlations using numerical data
for both isospin symmetric and asymmetric nuclear matter. In particular, we discuss
the impact of a finite isospin polarization on SRC. We also highlight the importance of
tensor components in a relatively high-momentum region.

In addition, several groups have recently tried to provide scaling arguments for the effect of SRCs in nuclei 
\cite{Frankfurt2009,Sargsian2011,Sargsian2012,Vanhalst2012,Alvioli2013}. 
In practice, one aims at identifying how these effects change with particle number and if they
are dominated by the number of particles or by the number of pairs.
From an infinite-matter perspective, if one stays at the one-body level, total and partial nucleon 
densities are the more obvious available quantities for such a scaling. 
In the following, we illustrate how the two are connected in the SCGF approach.
We base our analysis on the isospin asymmetry dependence of the dilute Fermi gas (DFG)
in the high-momentum limit (see the Appendix for details) 
\cite{Abrikosov1965,Sartor1980a,*Sartor1980b}.
We limit our discussion to the one-body momentum distribution, 
but note that the two-body distribution can be accessed in our formalism \cite{Dickhoff1999}.  

The variation	
of the results with the underlying nucleon-nucleon (NN) interaction provides an indication of the robustness 
of the physical effects at play. 
We choose four NN forces, with very different short-range and tensor components,
to explore the widest possible range.
We thereby aim to reduce model dependence and to identify trends 
within our many-body summation technique. The Argonne v18 (Av18) represents a 
phenomenological parametrization in real space, including 18 spin-isospin 
operators \cite{Wiringa1995}. 
It is a typical example of a particularly strong, but finite, short-range core. 
The charge-dependent Bonn (CD-Bonn) 
interaction is based on a meson-exchange picture and provides
a very accurate fit to the two-body sector \cite{Machleidt1995}. Its short-range core is softer than 
Av18, which yields less high-momentum components. We also use the Idaho 
next-to-next-to-next-to-leading-order (N3LO) chiral-perturbation-theory potential of Ref.~\cite{Entem2003}.
Its low-energy constants are fitted to reproduce the two-body sector with a 
large accuracy in a $\chi^2$ procedure (see Ref.~\cite{Ekstrom2013} for an alternative fitting 
strategy). Owing to its very nature, as an effective-field-theory potential, the interaction requires a 
momentum cut-off, which is chosen at $500$ MeV. In general, this NN force induces relatively 
few high-momentum components in the many-body wave function.
This is particularly true beyond the cut-off momentum, where the
potential is not expected to apply. These three NN interactions are phase-shift equivalent.  
In contrast, we also display a few results obtained with the Av4' reduction of Av18 
\cite{Wiringa2002}.
This potential does not reproduce experimental NN scattering phase shifts.
In particular, it lacks tensor terms, but it is built to reproduce the deuteron
binding energy. Av4' is therefore useful 
in assessing the importance of tensor correlations in the nuclear medium. 
A detailed analysis of the energy of nuclear and neutron matter with different versions 
of the Argonne potential has been recently reported in Ref.~\cite{Baldo2012}.

In the following, we neglect the effect of three-body forces. Recent calculations indicate 
that the isospin 
asymmetry dependence of SRCs is relatively insensitive to the presence of three-body
interactions, at least around saturation density \cite{Yin2013}. 
We note, however, that the usual BHF
approach to include effective two-body forces obtained from three-body interactions by a simple density folding
is not consistent with a diagrammatic expansion including antisymmetrized matrix elements 
\cite{Carbone2013b,*Carbone2013a}. 
Moreover, the density folding in isospin asymmetric matter has not been fully implemented \cite{Drischler2013}. 
Hence, a fully quantitative answer to this issue is still missing. 

The present study complements, from a microscopic perspective, a series of recent analyses of the 
impact of high-momentum components on isovector properties 
\cite{Xu2011,Vidana2011,Carbone2012}. 
Both schematic and purely microscopic calculations with different methods 
have identified the effect of SRCs in the kinetic component of the symmetry 
energy. In general, a correlated system has a larger kinetic energy than its uncorrelated
counterpart. The redistribution of one-body strength underlying this increase, however, is 
different for different asymmetries. 
In particular, the asymmetry dependence is milder in the correlated case, which leads to
a smaller kinetic symmetry energy. 
At low densities, this component can even become negative. 
Whether this reduction affects the observable bulk properties of neutron-star matter 
is still under discussion.  

\section{Formalism}

A particularly relevant question, in view of recent advances \cite{Sargsian2011,Sargsian2012}, is 
the dependence of SRCs with isospin asymmetry. In an arbitrarily isospin-polarized system, 
effects dominated by np pairs will have a different asymmetry dependence than those 
due to same-isospin pairs. This can potentially have observable implications on single-particle properties. 
Here we look at the problem from the perspective of isospin asymmetric nuclear matter 
\cite{Zuo1999}.
We will characterize isospin asymmetry using either the proton fraction, 
$x_p = \frac{\rho_p}{\rho_n+\rho_p}$, or the isospin asymmetry parameter, 
\begin{align}
\beta=\frac{\rho_n-\rho_p}{\rho_n+\rho_p} = 1-2x_p \, .
\label{eq:beta}
\end{align}
In practice, our calculations are performed at finite temperature to avoid instabilities
associated with pairing solutions \cite{Alm1993,Frick2005,Rios2009a}.
While a proper treatment of pairing can be implemented~\cite{Muther2005}, 
it is not relevant for the discussion of high-momentum components. 
In the following, all the results have been computed at a low temperature of $T=5$ MeV. 
Although not shown here explicitly, we have checked that our conclusions
are not seriously affected by working at finite temperature.
We note that the high-momentum region is dominated by NN correlations 
rather than thermal effects \cite{Rios2009a}, and hence it is insensitive to temperature.  

Correlations have a particularly clear manifestation in the single-particle momentum 
distribution \cite{Rios2009a},
\begin{align}
	n_\tau(k) = \langle a^\dagger_\tau(k) a_\tau(k) \rangle \, , 
\end{align}
where $a^\dagger(k)$ [$a(k)$] is the creation (destruction) operator of a single-particle
excitation with momentum $k$ and isospin index $\tau=n, p$. The average,
$\langle \cdot \rangle$, is taken over a complete thermal set of many-body states 
\cite{Dickhoff2008}.
Our aim is to quantify how one-body high-momentum components evolve with 
density and isospin asymmetry. 
To access these density and asymmetry dependencies, we perform 
SCGF calculations at arbitrary densities and isospin asymmetries \cite{Frick2005}. 
We provide here a concise explanation of the ladder SCGF method. 
A detailed description of the numerical techniques employed in its resolution can be found elsewhere
\cite{Frick2005,frickphd,*riosphd}.

The numerical simulations are obtained from a self-consistent summation of 
ladder diagrams in the two-body sector which accounts consistently for SRCs and tensor effects. 
This is achieved by means of an in-medium
Lippmann-Schwinger equation, which can be schematically represented as
\begin{align}
	\mathcal{T}_{\tau+\tau'} = V_{\tau+\tau'} + V_{\tau+\tau'} G_{II,\tau \tau'}^0 \mathcal{T}_{\tau+\tau'} \, .
\end{align}
The sums $\tau+\tau'$ stand for the three relevant channels for
 in-medium scattering in asymmetric matter, namely neutron-neutron (nn), proton-proton (pp) and np.
The isospin-dependent NN interaction, $V$, is summed to all orders to obtain a well-behaved 
in-medium effective interaction, or $\mathcal{T}-$matrix. We solve the Lippmann-Schwinger equation 
using matrix techniques with up to $J=4$ partial waves in all isospin channels. 

In-medium scattering is mediated by a lowest-order 
two-body propagator, $G_{II,\tau \tau'}^0$,
which accounts for both particle-particle and hole-hole correlations 
and incorporates off-shell effects completely. 
In asymmetric matter, the different isospin elements of this propagator are given by:
\begin{align}
\label{eq:g2}
&G_{II,\tau \tau'}^0(k,k'; \Omega) = \\
&\int \frac{\textrm{d} \w}{2 \pi} \frac{\textrm{d} \w'}{2 \pi} 
\frac{ \G^<_\tau(k,\w)  \G^<_{\tau'}(k',\w') - \G^>_\tau(k,\w)  \G^>_{\tau'}(k',\w')}{\Omega - \w - \w' + i \eta} \, . \nonumber
\end{align}
The two components of the single-particle propagator, $G^<_\tau$ and $G^>_\tau$, are
related by a Kubo-Martin-Schwinger relation \cite{Abrikosov1965}. They can also be linked to the single-particle
spectral function, $\A_\tau$,
\begin{align}
 \G^<_\tau(k,\omega) &= f_\tau(\omega) \A_\tau(k,\omega) \, , \\
\G^>_\tau(k,\omega) &= [1 - f_\tau(\omega)] \A_\tau(k,\omega) \, , 
\label{eq:gm}
\end{align}
using the Fermi-Dirac distribution of each species, 
$f_\tau(\omega) = [ 1 + e^{  \frac{ \omega - \mu_\tau}{T} } ]^{-1}$. 
The chemical potential, $\mu_\tau$, is found from the integral
of the momentum distribution, $n_\tau$. 
We normalize the distribution to unity:
\begin{align}
\int_0^\infty \!\!\!\!\! \textrm{d} k \, k^2 n_\tau (k) = 1 \, ,
\label{eq:momdis}
\end{align}
which allows for clear comparisons with a finite system like the deuteron \cite{Boeglin2011}. 
The momentum distribution is related to the single-particle propagator via a 
density-dependent factor:
\begin{align}
n_\tau(k) = \frac{2}{\pi^3 \rho_\tau} \int_{-\infty}^\infty \!\!\!\!\! \textrm{d} \omega \, G^<_\tau(k,\omega) \, .
\label{eq:momdis_gfun}
\end{align}
The combination of the two expressions, 
Eqs.~(\ref{eq:momdis}) and (\ref{eq:momdis_gfun}), 
provides a non-trivial connection between the 
partial density, $\rho_\tau$, and the chemical potential, $\mu_\tau$.

Self-consistency is imposed all the way through in our calculations. The 
single-particle propagators
that enter Eq.~(\ref{eq:g2}), for instance, are obtained from the $\mathcal{T}-$matrix itself. 
In the ladder approximation, 
this effective interaction defines the imaginary part of the (retarded) self-energy:
\begin{align}
\textrm{Im} \Sigma_\tau (k,\omega) &= 
\sum_{\tau'} \int \!\!\! \frac{\textrm{d}^3 k'}{ (2 \pi)^3} \int \! \frac{\textrm{d}\omega'}{2 \pi}  
\left[ f_{\tau'}(\omega') + b_{\tau,\tau'}(\omega + \omega')  \right] \nonumber \\
& \times \left\langle \bm{k} \bm{k}' \right| \textrm{Im} \, \mathcal{T}_{\tau+\tau'} ( \omega + \omega' )\left|  \bm{k} \bm{k}' \right\rangle_A 
\A_{\tau'}(k',\omega') \, ,
\label{eq:imself}
\end{align}
which  acquires  different contributions from equal and unequal isospin partners. 
The difference arises from both the 
fermionic, $f_\tau(\omega)$, and the bosonic, 
$b_{\tau,\tau'}(\Omega) = [ e^{  \frac{ \Omega - \mu_{\tau} -\mu_{\tau'}}{T}}  -1 ]^{-1}$,
phase-space factors. Furthermore, there is also an isospin dependence associated with the 
isospin-channel-dependent retarded $\mathcal{T}-$matrix, which is properly antisymmetrized in the relevant channels. 

The dispersive contribution to the real part of the self-energy can be obtained from
a dispersion relation 
\cite{Dickhoff2008}. 
The generalized, instantaneous Hartree-Fock contribution, 
\begin{align}
\Sigma^{HF}_\tau(k) = 
\sum_{\tau'} \int \!\!\! \frac{\textrm{d}^3 k'}{ (2 \pi)^3} \int \! \frac{\textrm{d}\omega'}{2 \pi}  
&\left\langle \bm{k} \bm{k}' \right| V_{\tau+\tau'} \left|  \bm{k} \bm{k}' \right\rangle_A \nonumber \\
& \times G^<_{\tau'}(k',\omega') \, ,
\label{eq:self_HF}
\end{align}
 must be included as well. 
We  consider  up to $J=8$ partial waves in the calculation of this contribution, which is relevant
for in-medium quasiparticle shifts. 
The solution of the Dyson equation generates the single-particle spectral function:
\begin{align}
\A_\tau(k,\w) = \frac{-2 \textrm{Im} \Sigma_\tau(k,\omega) }
{ [ \omega - \frac{k^2}{2m} - \textrm{Re} \Sigma_\tau(k,\omega) ]^2+ [ \textrm{Im} \Sigma_\tau(k,\omega) ]^2} \, .
\label{eq:asf}
\end{align}
Feeding this spectral function into Eqs.~(\ref{eq:g2})-(\ref{eq:gm}), 
one obtains a self-consistency loop that treats all particles in the same manner, 
providing feedback to the different ingredients of the calculations. 
This goes beyond  the scope of  other approaches in providing an asymmetry-dependent 
spectral function and hence a fully correlated description of the microscopic properties of 
the system.
In addition, macroscopic bulk properties can be accessed by means of sum-rules and
the Luttinger-Ward formalism \cite{Frick2005,Rios2009a,Carbone2012}.

\section{Density dependence of short-range correlations}

\begin{figure*}[t!]
  \begin{center}
    \includegraphics*[width=0.9\linewidth]{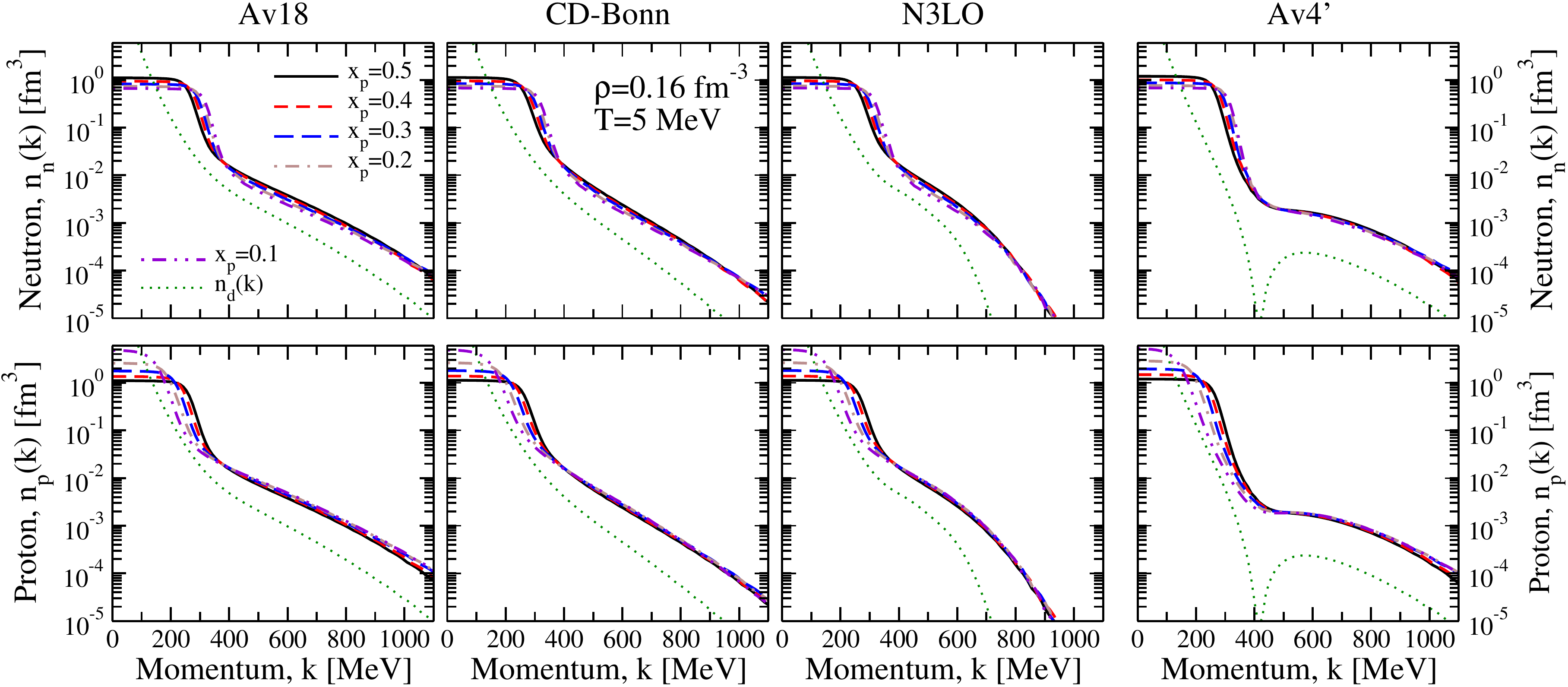}  
    \caption{(Color online) Momentum distribution for neutrons (top panels) and protons (bottom panels) in asymmetric nuclear matter. Different panels correspond to different NN interactions, from left to right: Argonne v18, CD-Bonn, N3LO, and Av4'. Lines represent data obtained at different proton fractions. We also show, for comparison, the momentum distribution of the deuteron (dotted line) associated with each interaction. }
    \label{fig:momdis_asy_deu}
  \end{center}
\end{figure*}

In our previous work, we have investigated the asymmetry dependence 
of the (depleted) low-momentum components of the one-body momentum distribution \cite{Rios2009a}. 
In that case, because the dependence on momentum 
below the Fermi surface is particularly mild \cite{Vonderfecht91,Mahaux1992},
one can focus on the single, lowest available momentum $k=0$. 
In contrast, any attempt to quantify effects associated with high 
momenta will necessarily depend on the definition of such ``high momenta." 
Here, we define high-momentum regions as ranging from $\sim 400$ to $\sim 850$ MeV. 
At the lower end, this is far enough from the Fermi surface to avoid effects associated to
its smoothing with temperature at different asymmetries\footnote{This momentum is also close to point at which
a pure $S-$wave deuteron would have a node in the momentum distribution; see right  panels of Fig.~\ref{fig:momdis_asy_deu}.}.
 At the high end, this range represents the limit at which momentum
distributions stop scaling, as we see next. 
Within high-momenta, we also define a ``tensor-dominated" momentum region, ranging from 
$k\sim 400$ MeV to $k \sim 550$ MeV (see below for details).    

We start the discussion by showing in Fig.~\ref{fig:momdis_asy_deu} the momentum distribution of 
neutrons (top panels) and protons (bottom panels) at different asymmetries for a fixed
total density of $\rho=0.16$ fm$^{-3}$ and a temperature of $T=5$ MeV. The 
different columns represent results obtained with different NN interactions. We focus our 
attention on the high-momentum region, well above the Fermi surface, 
and hence plot our results in a logarithmic scale. 
Overall, we note that the high-momentum components change very little when the isospin 
asymmetry is modified. This suggests that, when the momentum distribution is normalized
to unity, high-momentum components are basically determined by the total density of the system. 

We compare the results of asymmetric infinite matter to the corresponding deuteron 
momentum distribution, shown as the dotted line. 
Qualitatively, and as expected, the high-momentum components in
the momentum distribution of nuclear matter are qualitatively similar to those of the corresponding 
deuterons. 
For the phase-shift equivalent interactions, we identify an area, in the momentum region 
$400-550$ MeV, where the slopes of the two distributions are very much alike. 
In contrast, Av4' results (rightmost column), which lack tensor effects, show 
a node in the momentum distribution in this region. 
This confirms that the deuteron momentum distribution in this
area is dominated by tensor effects, as observed in previous studies \cite{Schiavilla2007}.
From now on, we call this the ``tensor-dominated" region. We quantify
the agreement between the two momentum distributions in the tensor-dominated region 
later. 

Looking at the high-momentum components in more detail, 
we observe that the intrinsic momentum cut off 
of N3LO shows up naturally in the momentum distribution (third column). 
The distribution displays a sharp decrease above $500$ MeV, absent in the other NN 
interactions. 
In particular, the deuteron momentum distribution data obtained in Ref.~\cite{Fomin2012} 
 is not reproduced for such
a soft interaction (or for an interaction with no tensor terms, such as Av4'). 
One can debate 
whether such high-momentum
components are physically motivated in NN interactions \cite{Bogner2007,Bogner2010}, but 
their presence in 
the many-body wave function seems unquestionable. 
While the short-range part of the NN interaction is not constrained by a fit to scattering data,
we note that all experiments indicate that no sharp drop associated 
with a cut off can be expected \cite{Boeglin2011,Fomin2012} .
We therefore adopt the interpretation that the short-range part of relative NN wave functions 
must be suppressed because the probability to find two intact nucleons at such distances must 
vanish on account of their intrinsic structure.
Which NN interaction is to be preferred can then be benchmarked by demanding an 
appropriate description of the modest $10 \, \%$ of nucleons with momenta not contained 
in the mean field, as measured, for instance, in Ref.~\cite{Rohe2004}.
Moreover, calculations with different short-range NN forces provide a
range of theoretical results. One can take this spread as a measure of the uncertainty
in the short-range description of NN interactions. 

We identify a small decreasing (increasing) trend in the neutron (proton)
high-momentum components as the system becomes more neutron rich, 
which we analyze in more detail in Sec.~\ref{sec:asymmetry}. 
Above $600$ MeV, the infinite matter results overshoot the deuteron ones. This happens for
all interactions, including Av4'. 
We take this as an indication of the larger importance of SRCs in a dense
system compared to the dilute deuteron bound state.  
In particular, the latter has a number density of $\rho \sim 2/(4\pi (5/3)^{3/2} r_d^3/3) \sim 0.02$ fm$^{-3}$, 
where $r_d$ is the deuteron charge radius. This density is well below saturation, where the results of
Fig.~\ref{fig:momdis_asy_deu} were obtained. 
We note that the SRCs overshooting  occurs  well above the tensor-dominated region and hence the two
regions should be distinguishable by applying momentum cuts. 

\begin{figure*}[t!]
  \begin{center}
    \includegraphics*[width=0.9\linewidth]{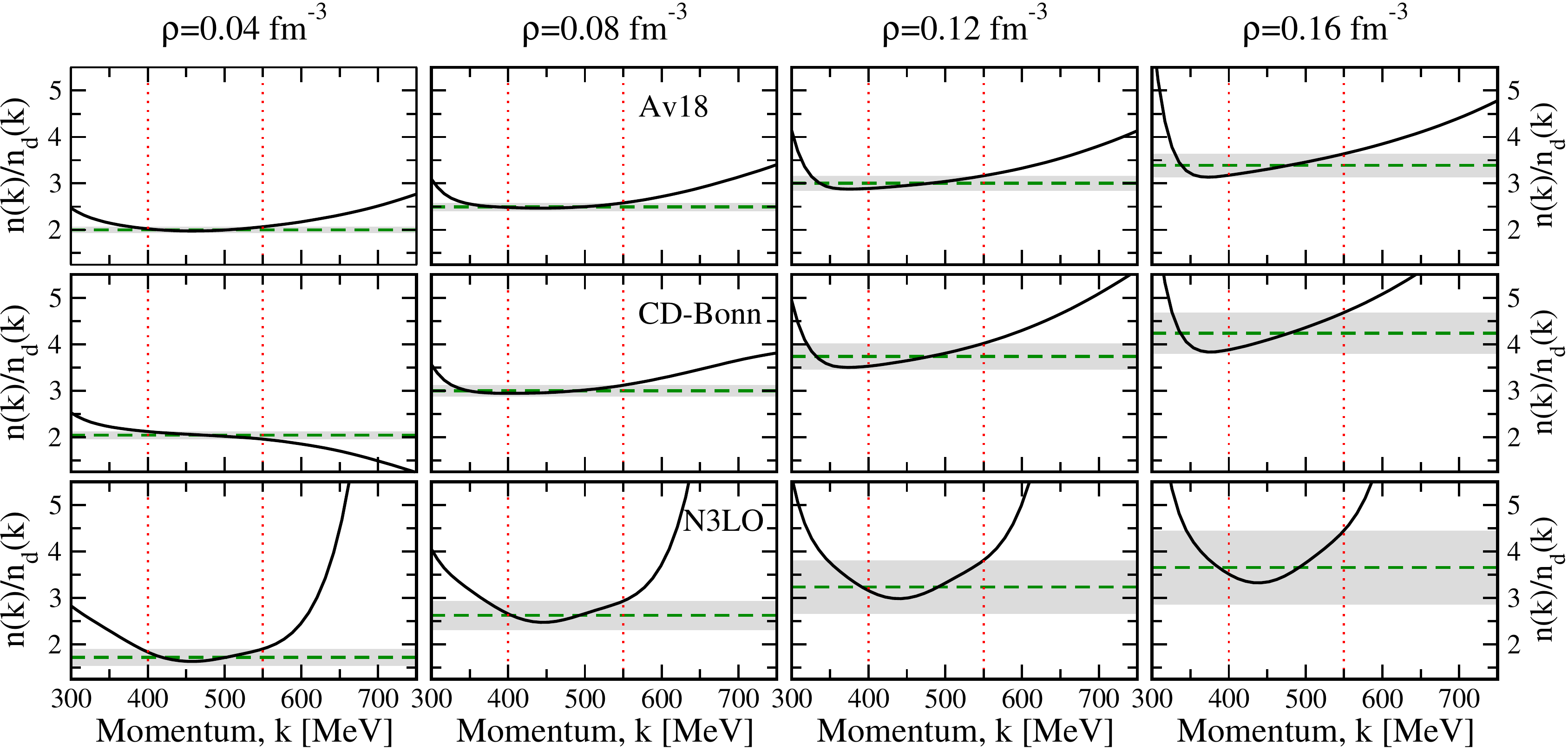}  
    \caption{(Color online) Ratio of the nucleon momentum distribution 
    to the corresponding deuteron distribution at high momenta. 
     Different columns correspond to different densities, from $\rho=0.04$ fm$^{-3}$ (left column) to 
     $\rho=0.16$ fm$^{-3}$ (right column) in equidistant steps. 
     Vertical panels show different NN interactions: Argonne v18 (top row), CD-Bonn (middle row) and
     N3LO (bottom row) panels. The dashed horizontal line and gray region represent, respectively, the average 
     and the error bar of the ratio in the tensor-dominated momentum region, depicted by vertical dotted lines. 
      }
    \label{fig:momdis_sym}
  \end{center}
\end{figure*}

One can quantify in more detail the similarity between $n_\tau(k)$ and $n_d(k)$, as well as the
gradual overshoot at high momenta, by looking at the ratio of the two momentum distribution.
This ratio is shown in continuous lines in Fig.~\ref{fig:momdis_sym} for different densities (columns) and three 
phase-shift equivalent interactions (rows). 
We draw three relevant observations from this figure.

First, there is a well-defined 
region in momentum where both distributions have similar momentum dependencies 
at subsaturation densities.
Qualitatively, one can take this plateau region as an indication of the dominance 
of tensor correlations. The latter are of a similar nature in the deuteron and in
the medium and hence lead to equivalent momentum dependencies in the distribution.
There are arguments in favor of such similarity, from either hard-scattering 
physics \cite{Amado1976} or factorization in the operator-product expansion \cite{Bogner2012}. 

Second, the ratio of the two distributions increases with density.
This is most clearly seen in the tensor-dominated region,
suggesting an increase in importance of tensor	like correlations with density. 
Because the relevant densities of  nuclei are fairly similar to those of the deuteron,
one should expect to see similar plateaus in finite systems \cite{Fomin2012}. 
Also, dense systems should have larger plateaus compared to dilute ones.

Third, the plateaus become less  pronounced  as the density increases. This suggests that 
the overshooting associated to SRCs occurs at lower momenta with increasing $\rho$. 
One expects the nuclear medium to induce
stronger correlations and hence lead to larger high-momentum tails as the density increases. 
This is particularly clear for N3LO (right panels), where
the effect is enhanced by the strong falloff of $n_d(k)$ above the cutoff momentum. 
Overall, this agrees with the idea that medium-induced SRCs
correlations become more relevant than tensor-based ones in the high-density region. 
The dominance of SRCs in the high-density region 
could have implications for neutron-star matter \cite{Frankfurt2009}.

Different measures have been proposed to quantify the importance of tensor correlations. 
The ratio of the momentum distribution of a given nuclear system to that of the deuteron 
is one of such quantities. 
We refer to this measure as $a_2$, following Ref.~\cite{Arrington2011}\footnote{Note that 
the same notation, namely $a_2$, has also been used to define other quantities 
such as cross section ratios \cite{Vanhalst2012,Arrington2012}. }.
More specifically, we define $a_2$ as the integrated average of the ratio in the tensor-dominated region:
\begin{align}
a_2 = \left< \frac{n(k)}{n_d(k)} \right>_{k=400-550 \text{MeV}} \, .
\label{eq:a2_av}
\end{align} 
This is related, but not necessarily equal, to the probability of a single nucleon to be part of a SRCs pair. 
Specifically, $a_2$ measures how many high-momentum nucleons there are in the distribution with
respect to those in the deuteron. 
The horizontal dashed lines in Fig.~\ref{fig:momdis_sym} correspond to the values of the averaged 
$a_2$ for the different NN interactions and densities. 
Our SCGF calculations suggest values of $a_2$ between $\sim 1$, at low densities, and $\sim 5$, at large densities.
The node in the deuteron momentum 
distribution for Av4' would lead to divergent values of $a_2$. This demonstrates once again the
relevance of tensor-induced correlations in this momentum region.
We note that changing the average integration region by $\pm 50$ MeV gives quantitatively similar results. 

For all three NN interactions, the value of $a_2$ increases with density.  The density dependence, 
however, is sensitive to the NN interaction under consideration. 
As already mentioned, the plateau in the
ratio of momentum distributions also becomes less and less prominent as the density increases. A clear
minimum is however present in all cases, which suggests that one can still characterize the ratio in the
tensor-dominated region using an average as in Eq.~(\ref{eq:a2_av}). 
To take into account any potential errors in the separation between tensor- and short-range-dominated regions,
we have assigned a conservative uncertainty to $a_2$.
This is represented in Fig.~\ref{fig:momdis_sym} by the gray bands, which have been 
obtained as the maximum deviation between the average and the function in the tensor-dominated momentum 
range. The shaded bands extend to momenta below the tensor-dominate region, 
cover the minimum in the ratio at all densities, 
and are wide enough to contain the numerical uncertainties associated with the solution of the SCGF equations.
In general, we find that the error in $a_2$ increases with density. 
We still obtain, however, a meaningful increasing trend as a 
function of density for all NN interactions, which we proceed to explore further.

\begin{figure}[t!]
  \begin{center}
    \includegraphics*[width=0.8\linewidth]{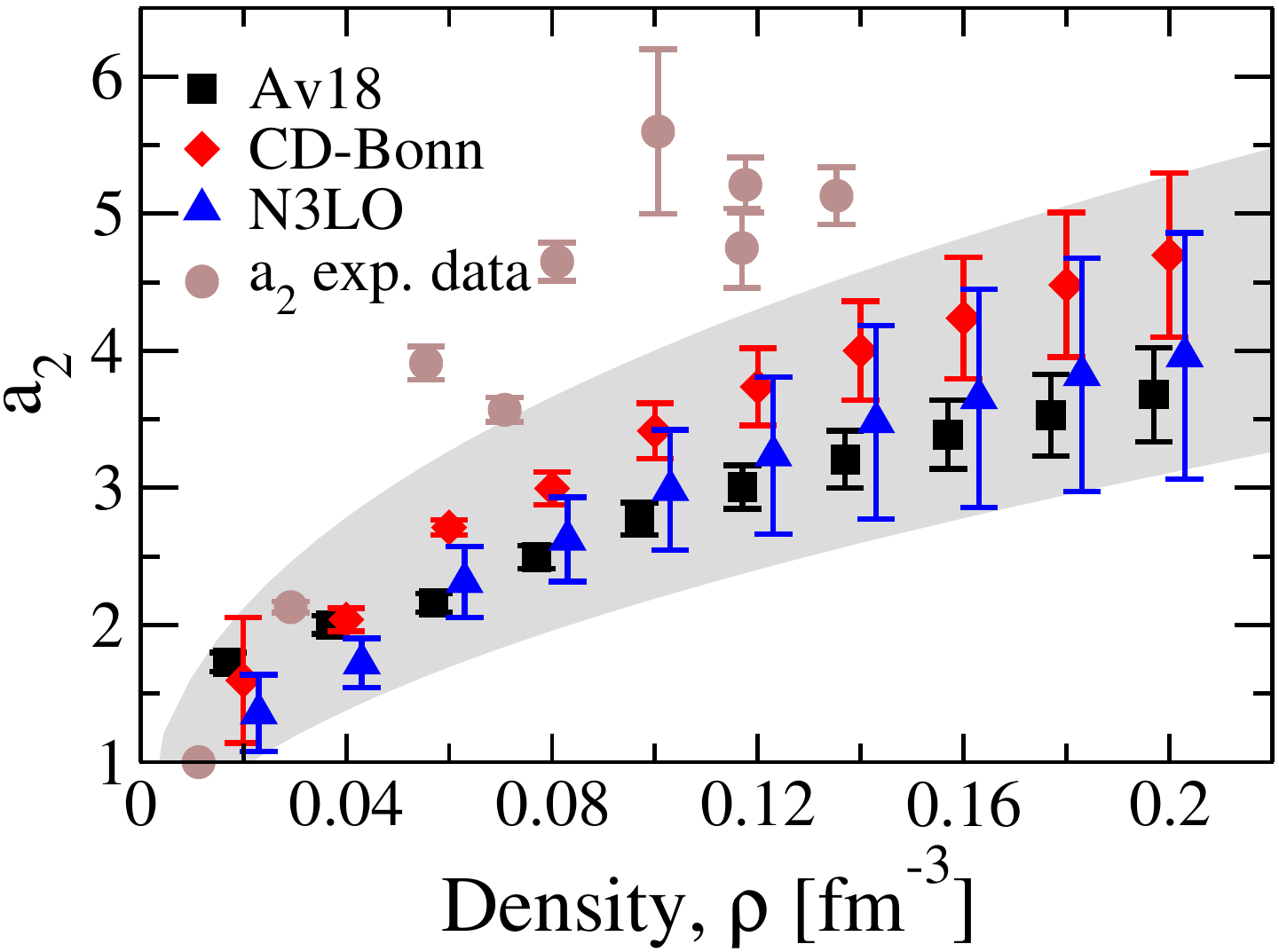}  
    \caption{(Color online) Density dependence of $a_2$, as defined in Eq.~(\ref{eq:a2_av}). 
    We show values obtained for three NN interactions, with a small horizontal offset for display purposes. 
    Experimental $a_2$ values from 
    Ref.~\cite{Arrington2012} are plotted as a function of a rescaled nuclear density (see text for details). 
    The shaded region corresponds to power laws with exponents $0.4-0.5$. 
    }
    \label{fig:a2den}
  \end{center}
\end{figure}

We show in Fig.~\ref{fig:a2den} the density dependence of $a_2$ obtained for the three relevant
NN forces. We note that, in all three cases, $a_2$ increases with density, saturating at high densities. 
We find modest differences between the potentials, which at half saturation fall in the
range $a_2 \sim 2-2.7$. 
Somewhat surprisingly, CD-Bonn seems to provide the largest $a_2$. 
In other words, compared to
the respective deuteron, CD-Bonn is a harder interaction than Av18.
As already mentioned, the error bars increase with density for all three 
interactions, particularly for N3LO, where SRCs are hardly present. 
A two-parameter fit of the type $a_2 = b_1 \rho^{b_2}$ provides a range of values 
$b_1=7-10$ and $b_2=0.4-0.5$. This yields the 
range depicted in the gray region in Fig.~\ref{fig:a2den}, which contains all data points and their error bars. 

We compare our values of $a_2$ to those extracted from a recent 
analysis of experimental data in Ref.~\cite{Arrington2012}.
Each experimental point in Fig.~\ref{fig:a2den} corresponds to a different isotope, between the deuteron and gold, 
where experimental $(e,e'p)$ data is available. These data have been reanalyzed recently in the context of the
connection between the EMC effect and SRCs \cite{Weinstein2011,Hen2012}.
Following this analysis, we have determined the number densities of the different isotopes
by assuming a spherical liquid drop with radius equal to the charge radius of Ref.~\cite{Angeli2013}.
To this density, we have also applied a $\frac{A-1}{A}$ correction to account for the excess density seen by the
knocked out nucleon, which propagates through a less dense environment. 
This should correspond to the so-called ``scaled nuclear density" of
Ref.~\cite{Arrington2012}. We note that at low densities (deuteron and triton, essentially), the experimental data 
agree well with our numerical estimates of $a_2$. At moderate and large densities, however, 
the experimental data is well above our results. 
Nevertheless, the data of Ref.~\cite{Arrington2012} are obtained as a ratio of deep inelastic cross sections rather than
as a ratio of momentum distributions. The connection between the observables and the momentum distributions is not
free of uncertainties. 
Let us stress, for instance, that our definition of $a_2$ does not account for the center-of-mass spreading of the np pair, 
a correction that can either reduce \cite{Arrington2012} or  increase \cite{Vanhalst2012} the values of $a_2$ according to different theoretical models.
A rescaling of the density or the theoretical values by a factor of $2$ would lead to a good overall
agreement between data and experiments. 

\section{Isospin asymmetry dependence of short-range correlations}
\label{sec:asymmetry}

\begin{figure*}[t!]
  \begin{center}
    \includegraphics*[width=0.75\linewidth]{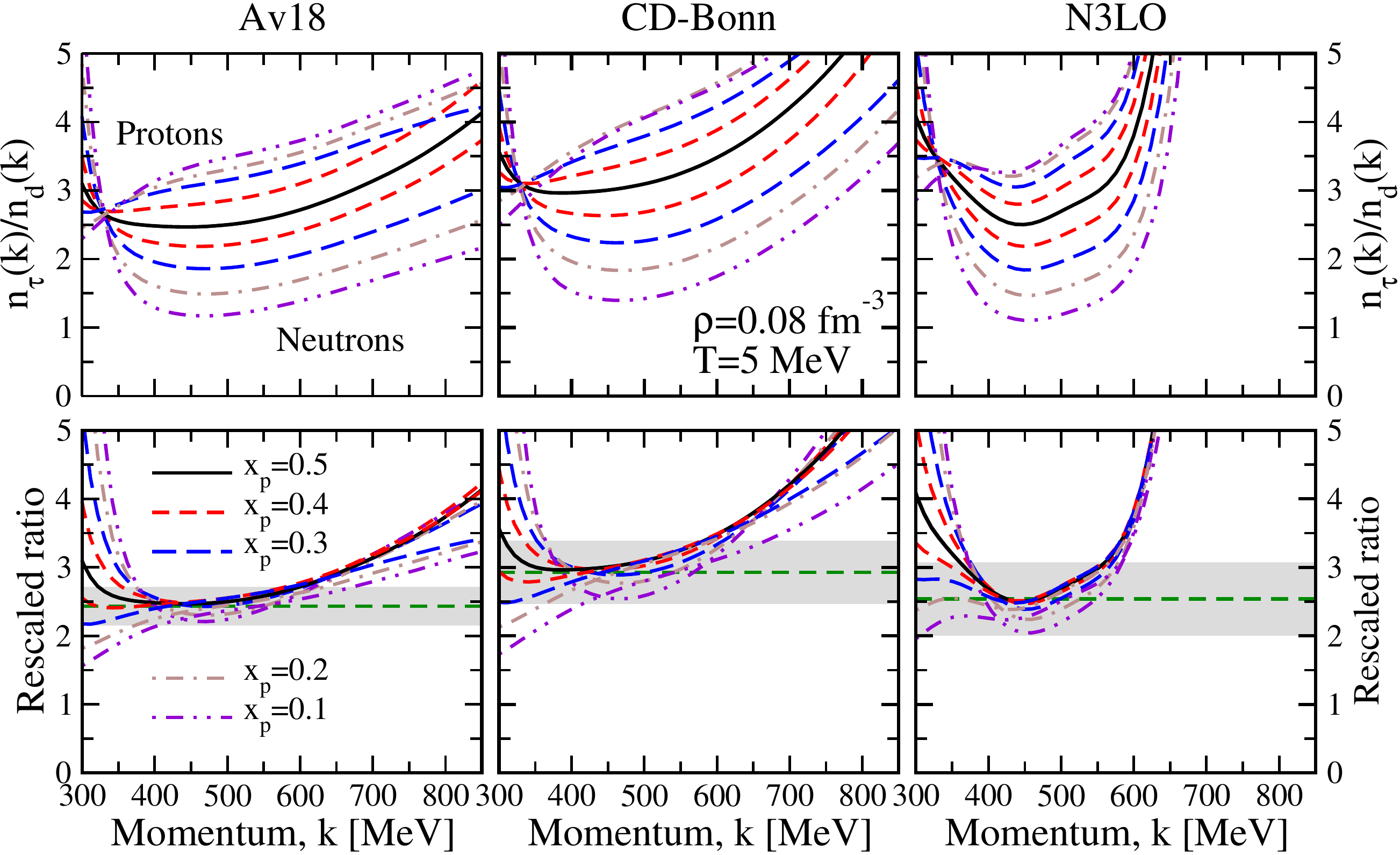}  
    \caption{(Color online) (Top panels) Ratio of the neutron and proton momentum distribution 
    to the corresponding deuteron distribution at high momenta. Different line styles correspond 
    to different isospin asymmetries. The three different panels correspond to thee 
    NN interactions, from left to right: Argonne v18, CD-Bonn and N3LO.
    (Bottom panels) Rescaled neutron and proton momentum distributions, as discussed in the text. 
    After rescaling, all results fall in a narrow band. Gray bands are an error band associated to the
    maximum deviation in the tensor-dominated region. }
    \label{fig:momdis_asy_deu_ratio}
  \end{center}
\end{figure*}

Our results indicate a qualitative resemblance of the one-body high-momentum components of the 
deuteron and symmetric nuclear matter. 
We now want to investigate whether a similar resemblance is found in 
isospin-asymmetric matter. Our final aim is to quantify how much tensor-dominated correlations
change with isospin asymmetry. We use the ratios of the  calculated  
momentum distributions and the deuteron distribution as tools to assess the isospin dependence
of tensorlike correlations. 
We show the ratio, for a fixed density of $\rho=0.08$ fm$^{-3}$ and different isospin asymmetries, 
in the top panels of Fig.~\ref{fig:momdis_asy_deu_ratio}. We have chosen this relatively low density 
because in the symmetric case it still shows a clear scaling.  
We provide results for the three realistic phase-shift equivalent potentials. 

The ratio of isospin-dependent momentum distributions reveals a noticeable dependence 
with isospin asymmetry that was not apparent in Fig.~\ref{fig:momdis_asy_deu}.
We find that the neutron distribution becomes less populated as the
system becomes more neutron rich. 
Conversely, we observe an increasing trend for the proton momentum distribution. Taking this population
as a measure of correlations, one can say that the minority species is more correlated. This agrees with
the conclusions we  drew when  investigating the depletion of low-momentum, isospin-asymmetric components
\cite{Rios2009a}.

For all but the highest asymmetries in the proton distributions, the ratios evolve linearly with isospin asymmetry. 
We justify this linear dependence in 
the Appendix, with a derivation based on the DFG formalism in the high-momentum limit. This indicates 
that the asymmetry dependence of the momentum distribution should be given by
\begin{align} 
n_\tau (k; \rho, \beta) = [1 \pm \gamma(\rho) \beta ] \times n_\tau (k; \rho, 0)  \, ,
\label{eq:scale}
\end{align}
where $\gamma(\rho)$ is a dimensionless, density-dependent 
parameter that quantifies the isovector strength of the effective interaction 
and $\beta$ is the isospin asymmetry as defined in Eq.~(\ref{eq:beta}).
More specifically, $\gamma$ should be associated with the isovector dependence of the effective interaction
in the tensor channel.
The latter parameter is negative, and therefore the plus (minus) sign corresponds to neutrons (protons). 
We have tested the validity of this asymmetry dependence for our numerical data 
at different densities and for a wide range of momenta. In particular, the asymmetry dependence
holds very well in the whole tensor-dominated, $k=400-550$ MeV range and for all asymmetries. 
Factorization in an isospin asymmetric medium can presumably lead to a more quantitative justification 
of this linear dependence with asymmetry \cite{Bogner2012}. 

The bottom panels of Fig.~\ref{fig:momdis_asy_deu_ratio} show rescaled ratios of the asymmetric matter and 
the deuteron momentum distributions. 
The rescaling is obtained by dividing the asymmetry-dependent momentum distributions by 
$ [1 \pm \gamma(\rho) \beta ]$. The optimal $\gamma$ parameter is fitted at each density. 
In practice, we minimize at each asymmetry the difference between the rescaled ratio and that 
of the symmetric case. We then average over asymmetries to find an optimal $\gamma$ for both
neutrons and protons at a given density. We provide error bars for this quantity in a similar fashion to  the analysis for 
$a_2$, finding the maximum deviations between the data and the averages. 
In principle, this procedure could be followed independently for both neutrons and protons, thus deriving
an isospin-dependent $\gamma$. Yet, we find a good agreement between the $\gamma$ parameters
obtained independently. We take this as a confirmation of the quality of our proposed
linear asymmetry dependence. 
We note that, after rescaling at a given density, the residual dependence on asymmetry in the tensor-dominated
momentum range is effectively removed. This is a rather general property, 
occurring at densities as high as $\rho \sim 0.24$ fm$^{-3}$. 

In contrast, Sargsian \cite{Sargsian2012} has suggested that a scaling of the type 
$n_\tau (k; \rho, \beta) \sim [1 \pm \beta]^{-\gamma} n_\tau (k; \rho, 0)$ 
with $\gamma \sim 1$ should be valid. 
We find that $\gamma=1$
does not provide a qualitatively good scaling of our data. In fact, it is not easy to find a power
that reproduces the ratios at all asymmetries.
We note, however, that the power-law 
scaling reduces to Eq.~(\ref{eq:scale}) in the small isospin-asymmetry limit. 
In particular, nearly symmetric systems like stable finite nuclei might be well reproduced by a power law
in asymmetry. 
Further work to validate this asymmetry dependence in finite nuclei would have to 
be explored with a different methodology \cite{Feldmeier2011,Vanhalst2012,Alvioli2013}.
We note, however, that extrapolations to neutron-star matter
require the explicit linear dependence of Eq.~(\ref{eq:scale}).

\begin{figure}[t!]
  \begin{center}
    \includegraphics*[width=0.75\linewidth]{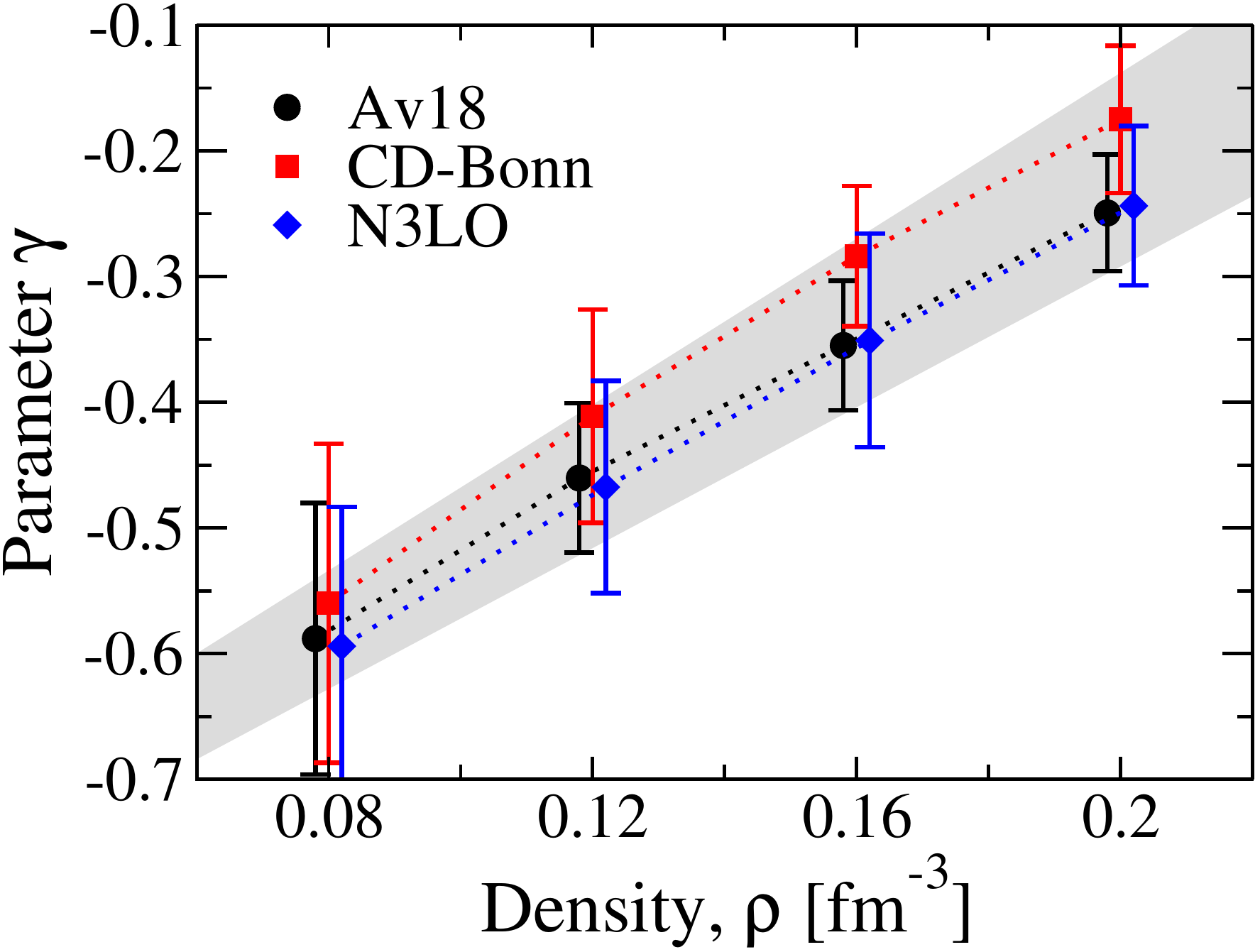}  
    \caption{(Color online) Density dependence of the $\gamma$ parameter for the three phase-shift equivalent 
    interactions considered in this work.
    Data for each NN force have a small horizontal offset for display purposes.  
    Error bars have been calculated as differences between the optimal neutron and proton $\gamma$ parameters. 
    The shaded region corresponds to linear fits with slopes $2.8-3.3$ fm$^3$. }
    \label{fig:gamma_rho}
  \end{center}
\end{figure}

We present in Fig.~\ref{fig:gamma_rho} 
the results of the averaged neutron and proton $\gamma$ computed at different densities. The error 
bars represent the maximum differences between the optimal fits of the neutron and proton 
distributions at different asymmetries. Again, this is a rather conservative estimate, 
generally larger than the fitting error of the individual $\gamma_\tau$.
In practice, the parameters are very close to each other for different phase-shift equivalent interactions, 
which suggests that this scaling is universal and independent of the short-range components. 
In other words, the isovector dependence of correlations in the tensor-dominated region
is very much restricted by phase-shift equivalence. 
Moreover, the density dependence of $\gamma$ is similar for
different interactions. In the range of densities considered, and with the present data, we find 
that the density dependence is almost linear. 
The gray band in Fig.~\ref{fig:gamma_rho} corresponds to a range of regressions 
with linear slopes $2.8-3.3$ fm$^3$ and origins between $-0.80$ and $-0.85$. 

Overall, our results suggest that the momentum distribution in asymmetric matter within the momentum range $400-550$ MeV is proportional to the deuteron momentum distribution.
Given our numerical SCGF data, we propose the following density and asymmetry dependence of $a_2$:
\begin{align}
a_{2,\tau}(\rho,\beta) = b_1 [1 \pm \gamma(\rho) \beta ]  \rho^{b_2} \, .
\label{eq:a_2}
\end{align}
We have confirmed the density dependence of the symmetric matter 
$a_2$ using our asymmetric momentum distribution data.
These are shown in a horizontal dashed line in the bottom panels of Fig.~\ref{fig:momdis_asy_deu_ratio}. 
The agreement with the $\beta=0$ case is good. The error bars are somewhat larger, owing to the 
additional uncertainties in the isospin-dependent nuclear-matter case. 

\begin{figure*}[t!]
  \begin{center}
    \includegraphics*[width=0.75\linewidth]{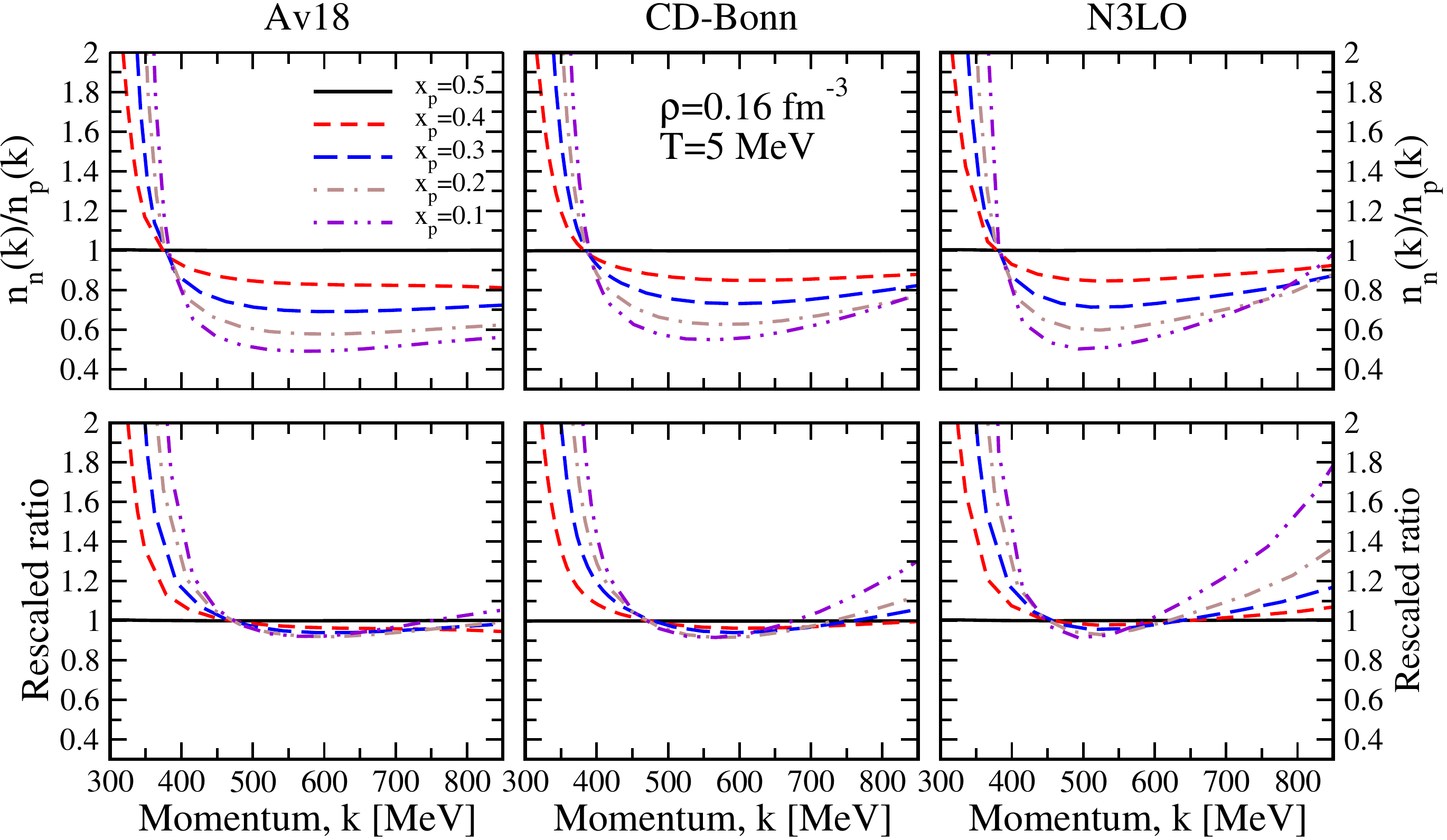}  
    \caption{(Color online) (Top panels) Ratio of the neutron to the proton momentum distribution for different proton fractions. (Bottom panels) Rescaled ratio of momentum distributions (see text for details). Different columns correspond to different NN interactions: Argonne v18 (left panels), CD-Bonn (middle panels), and N3LO (right panels). }
    \label{fig:mom_ratio}
  \end{center}
\end{figure*}

Up to this point, we have used our results to highlight a qualitative scaling of the high-momentum components with the local 
density of the nuclear medium. Further, one can discuss whether the 
momentum dependence itself is sensitive to the isospin asymmetry. The ratio
of momentum distributions,  $n_n(k)/n_p(k)$, provides useful  information on 
this issue. 
The top panels of Fig.~\ref{fig:mom_ratio} show the ratio at a fixed
total density for different isospin asymmetries in the tensor-dominated region. For symmetric
systems (solid line), the two distributions are the same at all momenta, as expected. 
When asymmetry is switched on ($x_p=0.4$), one observes a well-defined plateau in the whole 
high-momentum region, even beyond $k \sim 600$ MeV. 
The ratio is smaller than one, indicating that neutrons populate less the 
high-momentum components as compared to protons, as we have already seen.
This might be relevant for heavy nuclei, which have bulk asymmetries
of this order. Both momentum distributions have similar 
high-momentum  dependence, but with  proton populations about $15 \, \%$ larger than the ones for neutrons \cite{Sargsian2012}. 
At the largest asymmetries explored here, similar to those found in neutron stars, 
this percentage increases to $\sim 40 \, \%$.

At large asymmetries, the plateau is well defined for Av18 and CD-Bonn, 
which have a strong short-range core. The scaling is less defined in N3LO, owing to the
decay at momenta above $500$ MeV. In all cases, the departure from $1$ is stronger when 
asymmetry increases. The plateau in the momentum dependence also worsens as asymmetry
 increases. Again, this suggests a picture in which the asymmetry dependence of both momentum
distributions is attributed to the tensor effects, with protons populating more high-momentum components
as their fraction decreases. 
We note that the crossing around $400$ MeV is the same for all NN interactions. 

A further quantitative test of Eq.~(\ref{eq:scale}) can be obtained by rescaling the independent neutron
and proton momentum distributions with the same $\gamma$ and then taking their ratio. If all the isospin
asymmetry is correctly included in the scaling prefactor, all lines at different asymmetries should fall
within a narrow band. The bottom panels of Fig.~\ref{fig:mom_ratio} show the accuracy of this
rescaling procedure. We find that, after rescaling, all the asymmetry dependence is removed. The ratios
at different asymmetries fall within $10 \, \%$ of each other in the tensor-dominated region. 
Above $k\sim 600$ MeV, where SRCs dominate, we find differences for the larger asymmetry cases. 
This is particularly clear for N3LO. 
All in all, the figure suggests that the information on the momentum distribution 
in one asymmetric system might be enough to extrapolate to other asymmetric systems. 
A linear extrapolation using the universal $\gamma$ parameter is enough for these purposes. 
This procedure should work well for the tensor-dominated region, with larger uncertainties for
the short-range-dominated region. 

\section{Asymmetry dependence of integrated strength}

\begin{figure*}[t!]
  \begin{center}
    \includegraphics*[width=0.75\linewidth]{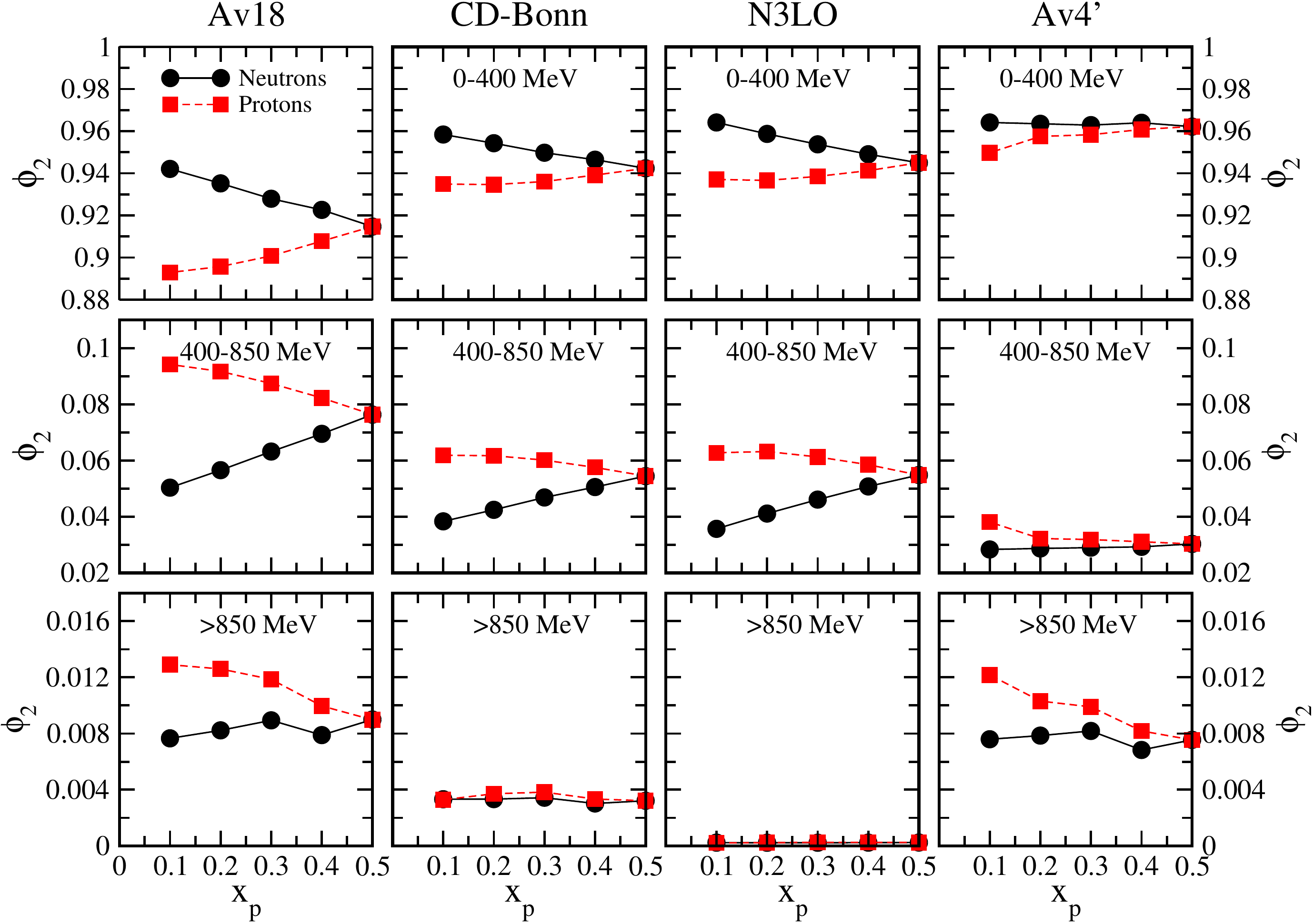}  
    \caption{(Color online) Integrated single-particle strength for neutrons (circles) and protons (squares) in three momentum regions: $k=0 - 400$ MeV (top panels), $k=400-850$ MeV (middle panels) and $k>850$ MeV (bottom panels). The four columns correspond to different NN interactions. Note the different vertical scales of the different rows. }
    \label{fig:strength}
  \end{center}
\end{figure*}

So far, we have paid particular attention to the momentum dependence of the distribution. In some cases, however,
one might be interested in looking at more global, or integrated, properties \cite{Sargsian2012}. 
We thus now proceed to analyze the integrated one-body strength in the high-momentum region. This
provides an insight into the mechanisms that populate high-momentum components and hence their 
importance beyond the independent-particle model. We quantify the population using the integrated 
strength,
\begin{align}
\phi_2(k_i,k_f) =  \frac{1}{\pi^2 \rho_\tau} \int_{k_i}^{k_f} \textrm{d} k \, k^{2} n_\tau(k) \, ,
\label{eq:strength}
\end{align}
in three different momentum regions. The low-momentum region, from $k_i=0$ to $k_f=400$ MeV,
includes depletion effects as well as the shifts in the Fermi momenta. In a free Fermi gas 
picture at $\rho=0.16$ fm$^{-3}$ and $T=5$ MeV, the thermal depletion in this region is 
very small at all asymmetries, in accordance
with the Pauli principle. The top panels of Fig.~\ref{fig:strength} show the asymmetry 
dependence of this region for both neutrons (circles) and protons (squares).
As expected, we find a depletion of states at low momenta, with a departure with respect to $1$. 
The observed isovector splitting is therefore necessarily induced by beyond-mean-field
correlations, with neutrons becoming more and more populated (that is, less 
correlated) as the system becomes more neutron-rich. This picture is in accordance with
our previous results \cite{Rios2009a}. 
Av4' data is less affected by isospin asymmetry compared to the data of other phase-shift 
equivalent potentials. We take this as yet another indication that tensor correlations are 
relevant for the isovector dependence of the momentum distribution. 

The middle panels of Fig.~\ref{fig:strength}  represent  the tensor-dominated region of interest for our study. 
In symmetric matter for phase-shift equivalent NN forces, this region contains between $6-8 \, \%$
of the single-particle strength. As asymmetry increases, however, high-momentum protons 
become more and more prominent. This is necessarily related to the decrease in the 
low-momentum region, because the overall strength is normalized.
 For a hard interaction such as Av18, and for fractions of 
$x_p \sim 0.1$, typical of neutron-star matter, up to $10 \, \%$ of the proton strength is in this 
high-momentum region. Conversely, neutrons populate less and less of the region, down to
$4-5 \, \%$, as the system becomes more neutron-rich. As we have already seen, 
the quantitative details 
depend on the short-range and tensor structure of the NN force under consideration. 
The qualitative picture, however, is extremely robust.
Protons become more prominent at high momenta as their concentration decreases. 
In this sense, one can say that protons are more 
correlated in a proton-poor environment. 
The limit $x_p \to 0$ would provide further insight, of relevance for polaron physics
\cite{Gubbels2013}. 

The bottom panels of Fig.~\ref{fig:strength} illustrate the remaining strength in the 
very high-momentum region, above $850$ MeV. 
Only interactions with a very strong short-range core, such as Av18, are able to promote
particles to this region, with typical populations of $\sim 1 \%$. We note that this happens also
for Av4', which indicates that this region is insensitive to the tensor structure of the force. 
The neutron-proton splitting here also becomes very small, if at all  relevant. 

Finally, we want to assess the importance of  high-momentum components in a bulk
 quantity  that depends on isospin asymmetry.
 The kinetic energy is particularly sensitive to such components 
 \cite{Carbone2012}, as it is an integral involving $n_\tau(k)$. 
We provide in Table~\ref{e_kin} the values of the neutron (proton) kinetic energy per neutron 
(proton) as a function of $x_p$. In symmetric matter, one observes the expected 
ordering, with the stronger short-range core potential, Av18, providing the largest values. 
Conversely, N3LO, the softest interaction, produces the lowest value.

\begin{table}[t!]
\begin{center}
	\begin{tabular}{c | c | c c c c c }
		\hline \hline
		$\phantom{a}$ & $x_p$ & $0.5$ &$0.4$ &$0.3$ &$0.2$ &$0.1$ \\ \hline
		\multirow{2}{*}{Av18} & $K_n$ & 41.8 & 42.3 & 44.2 & 44.8 & 45.6 \\
			& $K_p$ & 41.8 & 41.0 & 40.5 & 39.0 & 36.5 \\
		\hline
		\multirow{2}{*}{CD-Bonn} & $K_n$ & 35.0 & 36.5 & 38.4 & 39.8 & 41.2 \\
		& $K_p$ & 35.0 & 33.2 & 31.4 & 28.7 & 25.3 \\
		\hline
		\multirow{2}{*}{N3LO} & $K_n$ & 32.9 & 34.6 & 36.1 & 37.5 & 39.0 \\
		& $K_p$ & 32.9 & 31.1 & 28.8 & 26.2 & 22.7 \\
		\hline
	\end{tabular}
	\caption{Neutron (proton) kinetic energy per neutron (proton) for different proton fractions. All energies are in MeV and results for three NN interactions are shown. }
\label{e_kin}
\end{center}
\end{table}

As the isospin asymmetry increases, one observes a common trend: neutron kinetic energies 
increase, whereas proton energies decrease. This indicates that, even though protons 
dominate the high-momentum region, the growing contribution of neutrons in the 
low-momentum area still dominates the kinetic energy. 
The isovector splitting of kinetic energies is fairly symmetric and almost linear, except for
protons in the most asymmetric region, which we take as an indication of the appearance of 
low-momentum thermal effects. 
Curiously, even though Av18 provides
the largest isovector splittings in the strength, it provides the lowest splitting in kinetic 
energies. This has to do with the redistribution of spectral 
strength as a function of momentum, which
is weighted differently in the one-body strength than in the kinetic energy.

We also note that the total kinetic energy is the weighted
average of the proton and neutron components. Consequently, the neutron kinetic energy
dominates the result as $x_p$ decreases. The linear dependence of the individual 
energy fractions suggests that the total kinetic energy will be quadratic in isospin asymmetry. 
Overall, however, the isospin dependence is rather
mild, so that the kinetic component of the symmetry energy decreases with
respect to the free Fermi gas case \cite{Carbone2012}.  
This is the only effect that we are able to identify as correlation-driven. 
Our results do not seem to suggest that the tensor-dominated region is responsible
for any dramatic changes in total energy, at least around saturation density.

\section{Conclusions}

We have reviewed the density and 
isospin-asymmetry dependence of the high-momentum components
of the one-body momentum distribution. The latter were obtained at arbitrary asymmetries 
and different densities by means of SCGF ladder calculations. These are particularly well-suited
to study correlated momentum distributions, as they provide fully fragmented single-particle
propagators. To avoid the arbitrariness associated with
the short-range or tensor structure of the NN potential, we have performed calculations with a 
variety of interactions. These include phase-shift equivalent forces with strong and soft short-range cores
and with and without tensor terms.

Our  analysis shows that if the momentum distribution is normalized to one,
the high-momentum components of each NN force become almost universal and asymmetry-independent. 
Comparing with the associated deuteron of each interaction, we see that in-medium SRCs
tend to provide larger high-momentum components than the deuteron above $\sim 600$ MeV.
By taking a ratio of $n(k)$ to $n_d(k)$, however, we have been able to identify a region
of momentum, between $400$ and $550$ MeV, where
the two distributions have similar dependencies.
The comparison with a tensor-free interaction indicates that this region is dominated to a large
extent by the tensor components of the NN interactions.
The quality of the scaling of the two distributions decreases substantially with density because of the increasing importance of SRCs.  

The tensor-dominated region of $k\sim 400-550$ MeV is of interest, as it is expected that the momentum 
distribution here is rather universal \cite{Arrington2011}. Our calculations have demonstrated that
the ratio in this region can be characterized by a single parameter, 
which we refer to as $a_2$. This is related 
to the probability of finding short-range-correlated nucleons in the medium compared to the deuteron. The latter
increases with density and is somewhat dependent on the underlying NN interaction. We find that CD-Bonn 
provides the larger values of $a_2$, followed by N3LO and Av18. The uncertainty in this quantity
increases with density as well, but our predictions still fail to reproduce experimental values. 
This could either be attributable to an issue with pair counting or  with interpretation of the data. 
We have parametrized its density dependence in terms of a power law of exponent $\sim 0.4-0.5$. 
Interactions without tensor components, however, produce divergent results for $a_2$. 

We observe that isospin asymmetry shifts the in-medium momentum distributions to higher (lower)
values for protons (neutrons). This indicates a prevalence of high-momentum components 
of the minority species as the asymmetry increases, which confirms previous results. 
Using a DFG model as a guiding principle, we have verified that the residual
asymmetry dependence in the  high-momentum region is almost linear with asymmetry. The parameter
that drives this asymmetry dependence is negative, but approaches zero as the density increases. 
Moreover, we find that this parameter is independent of the NN interactions. This suggests that 
the isovector dependence of correlations is rather model independent, in agreement with 
previous results \cite{Rios2009a}.

Finally, we have turned our attention to integrated quantities, which also show a dependence on correlations and
asymmetry. Dividing the momentum distribution in different regions, we find that low-momentum neutrons 
are less correlated in neutron-rich matter. Conversely, the tensor-dominated region is predominantly populated
by protons as the system becomes more asymmetric. The isovector splittings are qualitatively similar for all
interactions, except for those missing tensor terms. Regarding the kinetic energy of each component, we 
have found an asymmetry dependence around saturation. However, the redistribution of 
strength is such that the neutron kinetic energy in neutron-rich systems is larger than in symmetric ones. In other 
words, even though neutrons populate more the low-momentum region as asymmetry increases, this change is not
as dramatic as that of the neutron fraction itself. We note that this asymmetry dependence has implications for
symmetry energy studies \cite{Carbone2012}.

Two-body observables can provide further insight into the interplay of high-momentum 
components, tensor correlations, and isospin asymmetry \cite{Dickhoff1999,Vanhalst2012,Alvioli2013}. 
These are beyond the scope of the  present work, but they can be generated in the SCGF framework.  
Access to such properties will provide further, much-needed quantitative understanding of 
SRCs in infinite and arbitrarily isospin-polarized nuclear systems. 

\begin{acknowledgments}

This work is partly supported by by the Consolider Ingenio 
2010 Programme CPAN CSD2007-00042, Grant No. FIS2011-24154 from MICINN 
(Spain) and Grant No. 2009GR-1289 from Generalitat de Catalunya 
(Spain); by STFC, through grants ST/I005528/1 and ST/J000051; by the U.S.
National Science Foundation under grants PHY-0968941 and PHY-1304242.

\end{acknowledgments}

\appendix
\section{Dilute Fermi gas}
The DFG model provides a  useful  perspective into the microscopic properties of correlated homogeneous 
fermionic systems. This model is based on the low-density Lee-Yang expansion for strongly interacting systems. 
Essentially, one 
recasts the many-body problem in terms of an effective, scattering matrix and
uses it in a perturbative expansion similar to Ref.~\cite{Galitski58}.
The low-energy scattering is characterized by a scattering length, $a$. Expressions for the momentum 
distribution of 
unpolarized systems were derived up to order $(k_F a)^2$ by Sartor and Mahaux \cite{Sartor1980a,*Sartor1980b}. 
We are not aware of any calculations for polarized systems using this model. 

For an unpolarized system of fermions with mass $m$, the momentum distribution in the region above the Fermi surface, $k>k_F$, is given by \cite{Abrikosov1965,Sartor1980a,*Sartor1980b}
 \begin{widetext}
\begin{align}
n(k) =  \frac{\nu(\nu-1)}{ (2 \pi^2)^3 } \frac{a^2}{ \rho m^2 }
\int_{k_1>k_F} \!\!\!\!\!\!\!\!\!\! d^3 {\bf k_1} 
\int_{k_2<k_F} \!\!\!\!\!\!\!\!\!\! d^3 {\bf k_2} 
\int_{k_3<k_F} \!\!\!\!\!\!\!\!\!\! d^3 {\bf k_3} 
\frac{\delta^{(3)} ( {\bf k} + {\bf k_1} - {\bf k_2} - {\bf k_3} )}{ \left[ \frac{ k^2 + k_1^2 - k_2^2 - k_3^2 }{2m} \right]^2 } \, ,
\label{eq:momdis_int}
\end{align}
where we have introduced the degeneracy, $\nu=4 (2)$ for symmetric (neutron) matter.
We only consider the $k \gg 1$ limit of this model, which is particularly easy to compute and  provides an indication of the asymmetry dependence. For an unpolarized system, the limit can be found by noticing that the external momentum becomes larger than any other scale in the system. Pulling it out of the integral, one finds
\begin{align}
n(k) \to n^\gg(k)=   \frac{\nu(\nu-1)}{ (2 \pi^2)^3 } \frac{a^2}{ \rho k^4}
\int_{k_2<k_F} \!\!\!\!\!\!\!\!\!\! d^3 {\bf k_2} 
\int_{k_3<k_F} \!\!\!\!\!\!\!\!\!\! d^3 {\bf k_3} 
= \frac{\nu-1}{\nu} \frac{ 8 a^2 \rho}{k^4} 
\label{eq:k4_momdis}
\end{align}
\end{widetext}
Similar power-law behaviors at high momenta, albeit with different exponents, are also obtained in hard scattering calculations \cite{Amado1976} or in the electron gas \cite{Bogner2012}. 

Let us now consider the isospin-polarized case. The neutron momentum distribution is affected by the scattering 
of both neutrons with 
neutrons (nn) and neutrons with protons (np). 
Each of these is mediated by its own scattering scattering length. 
In the high-momentum limit, the distribution has 
the same structure as Eq.~(\ref{eq:k4_momdis}) with $\nu=2$:
\begin{align}
n^\gg_{nn}(k) = 
\frac{ 4 a_{nn}^2 \rho_n}{k^4} \, .
\label{eq:nn_momdis}
\end{align}
The scattering between different particles is somewhat different, as the internal integrals in Eq.~(\ref{eq:momdis_int}) need to take into account the
different Fermi surfaces. With this in mind, the integral becomes
 \begin{widetext}
\begin{align}
n_{np}(k) =  \frac{\nu^2}{ (2 \pi^2)^3 } \frac{a_{np}^2}{ \rho_n m^2 }
\int_{k_1>k_F^p} \!\!\!\!\!\!\!\!\!\! d^3 {\bf k_1} 
\int_{k_2<k_F^p} \!\!\!\!\!\!\!\!\!\! d^3 {\bf k_2} 
\int_{k_3<k_F^n} \!\!\!\!\!\!\!\!\!\! d^3 {\bf k_3} 
\frac{\delta^{(3)} ( {\bf k} + {\bf k_1} - {\bf k_2} - {\bf k_3} )}{ \left[ \frac{ k^2 + k_1^2 - k_2^2 - k_3^2 }{2m} \right]^2 } \, .
\label{eq:momdis_int_np}
\end{align}
\end{widetext}
Note that this contribution has a different overall degeneracy factor, associated with the lack of exchange in this channel.
Taking again the $k \gg k^\tau_F$ limit, one finds
\begin{align}
n^\gg_{np}(k) = \frac{ 8 a_{np}^2 \rho_p}{k^4} \, .
\end{align}
Adding up the two contributions, we find the following expression for the high-momentum limit of the neutron momentum distribution:
\begin{align}
n^\gg_{n}(k; \rho,\beta) &= 
\frac{4}{k^4} \left[ a_{nn}^2 \rho_n + 2 a_{np}^2 \rho_p \right]  \, .
\end{align}
Introducing the total density and the isospin-asymmetry parameter [Eq.~(\ref{eq:beta})], one finds:
\begin{align}
n^\gg_{n}(k; \rho,\beta) = n^\gg(k; \rho,0) \times \left[ 1 + \gamma \beta \right] \, .
\label{eq:nggn}
\end{align}
Here we have introduced the high-momentum limit in the symmetric case,
\begin{align}
n^\gg(k; \rho,0) =  
\frac{2 \rho}{k^4} \left[ a_{nn}^2+ 2 a_{np}^2 \right]  \, .
\end{align}
An equivalent expression to Eq.~(\ref{eq:nggn}) with a minus sign holds for protons. 
The high momentum components of the momentum distribution
have the same $k^{-4}$ dependence as the unpolarized case \cite{Sartor1980a,*Sartor1980b}. 
Their overall prefactor, however, scales linearly with asymmetry. 
We have chosen to parametrize the isovector dependence of the scattering matrix
in terms of a dimensionless parameter:
\begin{align}
\gamma = \frac{a_{nn}^2 - 2 a_{np}^2}{ a_{nn}^2 + 2 a_{np}^2 }
\label{eq:gamma}
\end{align}
For the $^1S_0$ channel in NN scattering, we find $\gamma = -0.52$. 

Strictly speaking, however, the DFG model should not apply for nuclear systems around saturation densities.
Nuclear scattering lengths are large compared to $k_F$, so that the expansion should break down in this regime.
This signals the need to consider effective range effects as well as other many-body techniques. 
Moreover, the limit $k>>1$ is not necessarily of interest in our case, as the tensor-dominated region is not that far 
away from the Fermi surface itself around saturation.

Nevertheless, the DFG model is still a useful analytic tool that suggests scalings
of correlations. 
If we take the results obtained here as a guiding principle for the scaling of momentum distributions with 
isospin asymmetry, we find that the functional form agrees well with our numerical results. 
We show the density dependence of the effective $\gamma$ parameter in dense matter 
in Fig.~\ref{fig:gamma_rho}. Incidentally, the DFG $\gamma$ parameter has the 
same sign and order of magnitude than that in our dense matter calculations (see Fig.~\ref{fig:gamma_rho}).   
In nuclear matter, we interpret this parameter as a measure of the strength of in-medium 
isovector  correlation effects, rather than as a ratio of scattering lengths.

\bibliographystyle{apsrev_mod}
\bibliography{biblio}

\end{document}